\documentclass[twocolumn,prl,showpacs,amsmath,assymb,eufrak]{revtex4}
\usepackage[dvipdfm]{graphicx}
\usepackage{bm}
\usepackage[mathscr]{euscript}
\usepackage{stmaryrd}
\newcommand{\lavetrajs}{\langle\!\!\!\langle}
\newcommand{\ravetrajs}{\rangle\!\!\!\rangle}
\begin{document}

\title{
An expression for stationary distribution
 in nonequilibrium steady state
}

\author{Teruhisa S. Komatsu}%
\affiliation{Department of Arts and Sciences, University of Tokyo,
 Komaba 3-8-1, Meguro, Tokyo, Japan 153-8902}%
\author{Naoko Nakagawa}%
\affiliation{Department of Mathematical Sciences,
 Ibaraki University, Mito 310-8512, Japan}%

\date{\today}

\begin{abstract}
We study the nonequilibrium steady state realized in a general  
stochastic system attached to multiple heat baths and/or driven by an  
external force.  Starting from the detailed fluctuation theorem we  
derive concise and suggestive expressions for the corresponding  
stationary distribution which are correct up to the second  
order in thermodynamic forces.  The probability of a microstate $\eta$  
is proportional to $\exp[{\Phi}(\eta)]$ where
${\Phi}(\eta)=-\sum_k\beta_k\mathcal{E}_k(\eta)$ is the excess entropy  
change.  Here  $\mathcal{E}_k(\eta)$ is the difference between two kinds of conditioned  
path ensemble averages of excess heat transfer from the $k$-th heat bath  
whose inverse temperature is $\beta_k$.  Our expression may be verified  
experimentally in nonequilibrium states realized, for example, in  
mesoscopic systems.
\end{abstract}

\maketitle


A challenge in theoretical physics is to find an extension of the 
Boltzmann factor $e^{-\beta H}$ which works in nonequilibrium systems.  
The present work reports a promising result which definitely goes 
beyond the linear response theory.

To be precise our concern here is to obtain concise and
 accurate characterization of
 the stationary distribution
 associated with the nonequilibrium steady state (NESS)
 in a system with heat currents and/or under external drive.
If the system is indefinitely close to equilibrium, there is a compact  
expression for the NESS in terms of fluctuation of nonequilibrium  
currents (or entropy production) as is known in the linear response theory
\cite{Kubo}.
As for a general system far from equilibrium, exact but formal  
expressions (like (\ref{eq:RhoTauEsA}) below) for the NESS have been known and used as  
starting points of further studies of nonequilibrium physics
\cite{Zubarev,McLennan,KawasakiGunton}
.
As we shall discuss,
 however, such formal expressions as they are  
turn out to be far from useful or enlightening.

In the present Letter, we start from the detailed fluctuation theorem  
\cite{FT:Evans,FT:Gallavotti,FT:Kurchan,FT:Maes,FT:Crooks,Jarzynski}
and derive a novel general expression (\ref{eq:rhost}) for the NESS, which is  
correct up to the second order in thermodynamic forces.
The expression, which involves expectation values of excess entropy  
change in transient regimes, is neat and quite suggestive.
Since the expression can be rewritten as (\ref{eq:SexHeat}) using (excess) heat  
transfers, it may be verified in actual experiments.

We hope that the novel expression leads to a better understanding of  
general features of NESS, and can be an important step in the  
construction of future nonequilibrium statistical mechanics.


We consider a classical system attached to multiple heat baths and  
 driven by an external force.
We assume that the system has a well defined energy $H(\eta)$, where  
 $\eta$ denotes the microscopic state.
The change of $H(\eta)$ in any process is identical to the sum of the  
 total heat transfer into the system from the heat baths
 and the total work done by the external force.
We further assume $H(\eta)=H(\eta^*)$, where $\eta^*$ is the time  
 reversal of $\eta$, i.e., the state obtained
 by reversing the signs of all the momenta in $\eta$.

For simplicity, we here treat a stochastic process
 with discretized state space and time.
Extensions to continuum case are straightforward.
By $\eta_t$ we denote the (microscopic) state of the system at time $t$.
A path or a microscopic history is denoted as
 $\bm{\Gamma}_0^\tau=(\eta_0,\eta_{\Delta t},\cdots,\eta_\tau)$,
 where $\Delta t$ is the unit of time.
We mainly consider paths from $t=0$ to $t=\tau$
 and sometimes use simpler notation $\bm{\Gamma}$ instead of
 $\bm{\Gamma}_0^\tau$.
Time-reversal operation $\vartheta$ is defined as
$
{\vartheta}\bm{\Gamma}_0^\tau
	=(\eta_{\tau}^*,\eta_{\tau-\Delta t}^*,\cdots,\eta_{0}^*).
$
The path probability $P(\bm{\Gamma}_0^\tau)$
 that a path $\bm{\Gamma}_0^\tau$ is realized 
 is written as
$
P(\bm{\Gamma}_0^\tau)
=\rho_\mathrm{ini}(\eta_0)
\mathcal{T}(\bm{\Gamma}_0^\tau)
$
 where $\rho_\mathrm{ini}$ is the initial distribution
 and $\mathcal{T}$ is the transition probability associated
 with the history $\bm{\Gamma}_0^\tau$.

It has been established that, in a wide class of systems,
 the transition probabilities of a path $\bm{\Gamma}$
 and its time reversed path $\vartheta\bm{\Gamma}$
 satisfies the relation
\begin{subequations}
\label{eq:DFT}
\begin{equation}
\mathcal{T}(\bm{\Gamma})
/\mathcal{T}(\vartheta\bm{\Gamma})
=e^{{\hat S}(\bm{\Gamma})}
,
\label{eq:fluctuationtheorem}
\end{equation}
 with the total entropy production along $\bm{\Gamma}$ defined as
\begin{equation}
{\hat S}(\bm{\Gamma}):=-\bm{\beta}\cdot{\hat{\bm{Q}}}(\bm{\Gamma})
=-\sum_k\beta_k\hat Q_k(\bm{\Gamma}),
\label{eq:Sdef}
\end{equation}
\end{subequations}
 where $\hat Q_k(\bm{\Gamma})$ is
 the total heat transfer from the $k$-th heat bath
 (whose inverse temperature is $\beta_k$)
 to the system during the history $\bm{\Gamma}$.
Here we have written $\bm{\beta}=(\beta_1,\beta_2,\cdots)$ and
 $\hat{\bm{Q}}=(\hat{Q}_1,\hat{Q}_2,\cdots)$.
The relation (\ref{eq:DFT}) can be derived for
 Langevin models from the Onsager-Machlup path integrals
\cite{OnsagerMachlup},
 and also 
 for Hamiltonian systems \cite{Jarzynski}.
It is sometimes called the detailed fluctuation theorem
\footnote{
	  This type of relation might be
	  called as ``local detailed balance condition'',
	  or ``microscopic reversibility''.
	  It is noted that Eqs. (\ref{eq:DFT}) holds even
	  under external drive.
	 }.

In the present work we study the NESS maintained by heat currents
 and/or a static external force $f_\mathrm{ext}$
 in a general system satisfying the symmetry (\ref{eq:DFT}).
By taking as a reference the equilibrium state
 where all the heat baths have the same inverse temperature
 $\beta^\mathrm{eq}$ and $f_\mathrm{ext}=0$,
 a NESS is specified by
 the thermodynamic forces
\begin{equation}
\bm{\epsilon}:=(\beta_1-\beta^\mathrm{eq}
 ,\beta_2-\beta^\mathrm{eq}
 ,\cdots,\beta^\mathrm{eq}f_\mathrm{ext}).
\end{equation}
%
By our assumption,
 the change of the energy during $\bm{\Gamma}$,
 $H(\eta_\tau)-H(\eta_0)$,
 is equal to $\sum_k\hat{Q}_k(\bm{\Gamma})+\hat{W}(\bm{\Gamma})$,
 where $\hat{W}(\bm{\Gamma})$ is the external work done to the system
 during $\bm{\Gamma}$.
We assume $\hat{W}$ is written as
 $\hat{W}=f_\mathrm{ext}\cdot\hat{R}_\mathrm{e}$,
 where $\hat{R}_\mathrm{e}$ is displacement
 conjugate to $f_\mathrm{ext}$.

In the following, path dependent observables
 are denoted as ${\hat X}(\bm{\Gamma}_0^\tau)$,
 ${\hat X}(\bm{\Gamma})$, or ${\hat X}$.
Path ensemble average of ${\hat X}$ is denoted by
$
\lavetrajs\hat{X}\ravetrajs:=\sum_{\bm{\Gamma}}\hat{X}(\bm{\Gamma})P(\bm{\Gamma}).
$
We define the ``conditioned path ensemble average'' (CPEA) as
\begin{equation}
\lavetrajs {\hat X} \ravetrajs_{\eta_t=\eta}
:=
\lavetrajs \delta_{\eta_t,\eta} {\hat X} \ravetrajs
/ \lavetrajs \delta_{\eta_t,\eta} \ravetrajs
,
\end{equation}
 where the subscript ``$\eta_t=\eta$''
 specifies the condition.

Let $\mathcal{T}(\bm{\Gamma};\bm{\epsilon})$
 denote the transition probability
 under the thermodynamic forces $\bm{\epsilon}$.
We consider path probabilities
\begin{equation}
\begin{array}{rcl}
P^{\mathrm{st}}(\bm{\Gamma}_0^\tau)
&:=&\rho_\mathrm{st}(\eta_0)
\mathcal{T}(\bm{\Gamma}_0^\tau;\bm{\epsilon}),\\
P^{\mathrm{eq}}(\bm{\Gamma}_0^\tau)
&:=&\rho_\mathrm{can}(\eta_0)
\mathcal{T}(\bm{\Gamma}_0^\tau;\bm{0}),\\
P^{\mathrm{es}}(\bm{\Gamma}_0^\tau)
&:=&\rho_\mathrm{can}(\eta_0)
\mathcal{T}(\bm{\Gamma}_0^\tau;\bm{\epsilon}),
\end{array}
\label{eq:Pdef}
\end{equation}
 where $\rho_\mathrm{st}(\eta)$ is the stationary distribution in the NESS and
 $\rho_\mathrm{can}(\eta)
 :=\exp(-\beta^\mathrm{eq}H(\eta))/Z(\beta^\mathrm{eq})$
 is the canonical distribution at the equilibrium.
The path ensemble averages defined by
 $P^{\mathrm{st}},P^{\mathrm{eq}}$ and $P^{\mathrm{es}}$
 are denoted as
 $\lavetrajs\cdot\ravetrajs^{\mathrm{st}}$,
 $\lavetrajs\cdot\ravetrajs^{\mathrm{eq}}$ and
 $\lavetrajs\cdot\ravetrajs^{\mathrm{es}}$, respectively. 
Similarly, the CPEAs defined with
 $P^{\mathrm{st}},P^{\mathrm{eq}}$ and $P^{\mathrm{es}}$
 are denoted by
 $\lavetrajs\cdot\ravetrajs^{\mathrm{st}}_{\eta_t=\eta}$,
 $\lavetrajs\cdot\ravetrajs^{\mathrm{eq}}_{\eta_t=\eta}$ and
 $\lavetrajs\cdot\ravetrajs^{\mathrm{es}}_{\eta_t=\eta}$.

Eq. (\ref{eq:fluctuationtheorem}) and 
 the third equation in (\ref{eq:Pdef}) lead to
\begin{equation}
P^{\mathrm{es}}(\bm{\Gamma})
/P^{\mathrm{es}}(\vartheta\bm{\Gamma})
=e^{{\hat S}_\mathrm{I}(\bm{\Gamma})}
,
\label{eq:fluctuationtheoremSI}
\end{equation}
 where ${\hat S}_\mathrm{I}$ is defined as
\begin{equation}
\begin{array}{rcl}
{\hat S}_\mathrm{I}(\bm{\Gamma})
&:=&
{\hat S}(\bm{\Gamma})
+\beta^\mathrm{eq} [ H(\eta_\tau)-H(\eta_0)]
\\&=&
 - \Delta\bm{\beta}\cdot{\hat{\bm{Q}}}(\bm{\Gamma})
 + \beta^\mathrm{eq}\hat{W}(\bm{\Gamma}).
\end{array}
\label{eq:SI}
\end{equation}
Here we set $\Delta\bm{\beta}:=
(\beta_1-\beta^\mathrm{eq},\beta_2-\beta^\mathrm{eq},\cdots)$
 and use $\hat{W}=H(\eta_\tau)-H(\eta_0)-\sum_k\hat{Q}_k$.
Using thermodynamic forces $\bm{\epsilon}$,
 the quantity ${\hat S}_\mathrm{I}$ is written with
 $\hat{\bm{R}}:=(\hat{\bm{Q}},\hat{R}_\mathrm{e})$ as
\begin{equation}
{\hat{S}}_\mathrm{I}(\bm{\Gamma})
=\bm{\epsilon}\cdot{\hat{\bm{R}}}(\bm{\Gamma}),
\label{eq:SIepsilon}
\end{equation}
which implies that
 $\hat{S}_\mathrm{I}$ is $O(|\bm{\epsilon}|)$ and 
 $\hat{S}_\mathrm{I}=0$ at the equilibrium.

Using Eqs. (\ref{eq:fluctuationtheoremSI})
 and the symmetry
 ${\hat S}_\mathrm{I}(\bm{\Gamma})
 =-{\hat S}_\mathrm{I}(\vartheta\bm{\Gamma})$,
 we obtain the relation
\begin{equation}
\lavetrajs{\hat X(\bm{\Gamma})}\ravetrajs^{\mathrm{es}}
=\lavetrajs
 {\hat X}(\vartheta\bm{\Gamma})\,
 e^{-{\hat S}_\mathrm{I}(\bm{\Gamma})}
\ravetrajs^{\mathrm{es}}
.
\label{eq:XESEnsembleAverage}
\end{equation}
By letting ${\hat X}=\delta_{\eta_\tau,\eta}$
 in Eq. (\ref{eq:XESEnsembleAverage}) we get a well-known
 formal expression
\begin{equation}
\rho_\tau(\eta)=\rho_\mathrm{can}(\eta^*)
\lavetrajs
e^{-{\hat S}_\mathrm{I}(\bm{\Gamma})}
\ravetrajs^{\mathrm{es}}_{\eta_0=\eta^*},
\label{eq:RhoTauEsA}
\end{equation}
where $\rho_\tau(\eta)$ is the distribution at time $\tau$
 evolved under the thermodynamic forces $\bm{\epsilon}$
 from the initial canonical distribution.
For sufficiently large $\tau$, 
 the distribution $\rho_\tau(\eta)$ should converge to
 the stationary distribution in the NESS, $\rho_\mathrm{st}(\eta)$.
To obtain Eq. (\ref{eq:RhoTauEsA}), we used the trivial relations 
 $\rho_\tau(\eta)=\lavetrajs \delta_{\eta_\tau,\eta} \ravetrajs^{\mathrm{es}}$ and
 $\rho_\mathrm{can}(\eta^*)=\lavetrajs \delta_{\eta_0,\eta^*} \ravetrajs^{\mathrm{es}}$.

The expression (\ref{eq:RhoTauEsA}) is an example of exact but formal expressions of  
NESS, which have been well-known and widely used \cite{Zubarev,McLennan,KawasakiGunton}.
Although (\ref{eq:RhoTauEsA}) is exact, it is hardly useful (as it is) because it  
involves the average over the whole (long) history.
Moreover since the quantity $\hat{S}_\mathrm{I}(\Gamma)$ typically diverges linearly  
in time, the quantity $\exp[-\hat{S}_\mathrm{I}(\Gamma)]$ should exhibit wild  
fluctuation.

We note in passing that when the system is indefinitely close to  
equilibrium, one may approximate (\ref{eq:RhoTauEsA}) as
\begin{equation}
\begin{array}{c}
\rho_\tau(\eta)
\simeq
\rho_\mathrm{can}(\eta^*)\,
\exp
\left[
{-\lavetrajs \hat{S}_\mathrm{I}(\bm{\Gamma})
\ravetrajs^{\mathrm{eq}}_{\eta_0=\eta^*}}
\right],
\\=
\rho_\mathrm{can}(\eta^*)\,
\exp
\left[
{-\bm{\epsilon}\cdot
\lavetrajs \hat{\bm{R}}(\bm{\Gamma})
\ravetrajs^{\mathrm{eq}}_{\eta_0=\eta^*}}
\right]
,
\end{array}
\label{eq:linearPDF}
\end{equation}
which can be used as a starting point of the linear response theory
\cite{Kubo,HayashiSasa}
.

We shall now derive our expressions (\ref{eq:rhost}), (\ref{eq:SexHeat})
 for NESS, which are almost as concise as (\ref{eq:linearPDF})
 but take properly into account nonlinear effects.
By letting ${\hat X}=\delta_{\eta_\tau,\eta}e^{-\hat{S}_\mathrm{I}/2}$
 in Eq. (\ref{eq:XESEnsembleAverage}), we obtain
\begin{equation}
\rho_\tau(\eta)=
{\rho_\mathrm{can}(\eta^*)}
\lavetrajs
e^{-{\hat S}_\mathrm{I}/2}
\ravetrajs^{\mathrm{es}}_{\eta_0=\eta^*}
\left/
\lavetrajs
e^{-{\hat S}_\mathrm{I}/2}
\ravetrajs^{\mathrm{es}}_{\eta_\tau=\eta}
\right.
.
\label{eq:RhoTauEsC}
\end{equation}
Let us consider the standard cummulant expansion
\begin{equation}
\begin{array}{c}
\ln
\lavetrajs
e^{- {{\hat S}_\mathrm{I}}/{2}}
\ravetrajs^{\mathrm{es}}_{\eta_t=\eta}
=
-\frac{1}{2}
\lavetrajs
{\hat S}_\mathrm{I}
\ravetrajs^{\mathrm{es}}_{\eta_t=\eta}
+\frac{1}{8}
\lavetrajs
{\hat S}_\mathrm{I};
{\hat S}_\mathrm{I}
\ravetrajs^{\mathrm{es}}_{\eta_t=\eta}
+O(|\bm{\epsilon}|^3)
,
\\
=
-
\frac{1}{2}
\lavetrajs
{\hat S}_\mathrm{I}
\ravetrajs^{\mathrm{es}}_{\eta_t=\eta}
+\frac{1}{8}
\lavetrajs
{\hat S}_\mathrm{I};
{\hat S}_\mathrm{I}
\ravetrajs^{\mathrm{eq}}_{\eta_t=\eta}
+O(|\bm{\epsilon}|^3)
,
\end{array}
\label{eq:cummulantexp}
\end{equation}
where we generally write
$
\lavetrajs A;B \ravetrajs
:=
\lavetrajs AB \ravetrajs
-
\lavetrajs A \ravetrajs
\lavetrajs B \ravetrajs
.
$
To get the second line in (\ref{eq:cummulantexp})
 we noted that 
$
\lavetrajs
{{\hat S}_\mathrm{I}};
{{\hat S}_\mathrm{I}}
\ravetrajs^{\mathrm{eq}}_{\eta_t=\eta}
$
 is already a quantity of $O(|\bm{\epsilon}|^2)$,
 and the nonequilibrium correction only gives
 a contribution of $O(|\bm{\epsilon}|^3)$.
Substituting Eq. (\ref{eq:cummulantexp}) into Eq. (\ref{eq:RhoTauEsC}),
 we obtain
\begin{equation}
\begin{array}{l}
\rho_\tau(\eta)=\rho_\mathrm{can}(\eta)\times
\\
~~\exp\left[
\frac{1}{2}
\left(
\lavetrajs
{\hat S}_\mathrm{I}
\ravetrajs^{\mathrm{es}}_{\eta_\tau=\eta}
-
\lavetrajs
{\hat S}_\mathrm{I}
\ravetrajs^{\mathrm{es}}_{\eta_0=\eta^*}
\right)
+O(|\bm{\epsilon}|^3)
\right]
,
\end{array}
\label{eq:LnRhoTauSI}
\end{equation}
 where the contributions from the second cummulants
 in $O(|\bm{\epsilon}|^2)$ have canceled out,
 because
$
\lavetrajs
{{\hat S}_\mathrm{I}};
{{\hat S}_\mathrm{I}}
\ravetrajs^{\mathrm{eq}}_{\eta_0=\eta^*}
=
\lavetrajs
{{\hat S}_\mathrm{I}};
{{\hat S}_\mathrm{I}}
\ravetrajs^{\mathrm{eq}}_{\eta_\tau=\eta}.
$
This follows from the relation
 in equilibrium,
$
\lavetrajs
{\hat X(\bm{\Gamma})}
\ravetrajs^{\mathrm{eq}}_{\eta_0=\eta^*}
=
\lavetrajs
{\hat X(\vartheta\bm{\Gamma})}
\ravetrajs^{\mathrm{eq}}_{\eta_\tau=\eta}
.
$

Considering Eq. (\ref{eq:LnRhoTauSI}) for sufficiently large $\tau$,
 we can further rewrite it into the form
 with path ensemble averages in NESS,
 while Eq. (\ref{eq:LnRhoTauSI}) is written as that
 in transient processes from equilibrium to NESS.
In the limit $\tau\to\infty$,
 the two CPEAs in the exponent on the r.h.s. of Eq. (\ref{eq:LnRhoTauSI})
 diverge but these divergences cancel with each other
 because the both CPEAs diverge with the same rate $\sigma\tau$
 where $\sigma$ is the entropy production rate in the NESS.

Substituting Eq. (\ref{eq:SI}) into Eq. (\ref{eq:LnRhoTauSI})
 and taking the limit of large $\tau$, we thus have
\begin{equation}
\begin{array}{l}
\rho_\mathrm{st}(\eta)
\propto
\exp\left[
\frac{1}{2}
\left(
\lavetrajs
{\hat S}
\ravetrajs^{\mathrm{es}}_{\eta_\tau=\eta}
-
\lavetrajs
{\hat S}
\ravetrajs^{\mathrm{es}}_{\eta_0=\eta^*}
\right)
+O(|\bm{\epsilon}|^3)
\right],\\
~~
=
\exp\left[
\frac{1}{2}
\left(
\lavetrajs
{\hat S}_\mathrm{ex}
\ravetrajs^{\mathrm{es}}_{\eta_\tau=\eta}
-
\lavetrajs
{\hat S}_\mathrm{ex}
\ravetrajs^{\mathrm{es}}_{\eta_0=\eta^*}
\right)
+O(|\bm{\epsilon}|^3)
\right]
,
\label{eq:rhoTauSex}
\end{array}
\end{equation}
 where the excess entropy change ${\hat S}_\mathrm{ex}$ 
 is defined as
\begin{equation}
{\hat S}_\mathrm{ex}:={\hat S}-\tau\sigma
,\;\;\;\;
\sigma:=\lim_{\tau\rightarrow\infty}\frac{1}{\tau}
\lavetrajs{\hat S}\ravetrajs^{\mathrm{st}}.
\end{equation}
Here $\sigma$ is
 the entropy production rate in the NESS.
To get Eq. (\ref{eq:rhoTauSex}),
 we omitted $\eta$-independent parts,
 noted that the system loses the memory of the initial condition
 for sufficiently large $\tau$, i.e.
$
\lim_{\tau\to\infty}
\lavetrajs
H(\eta_\tau)
\ravetrajs^{\mathrm{es}}_{\eta_0=\eta^*}
=
\lavetrajs
H(\eta_\tau)
\ravetrajs^{\mathrm{es}},
$
 and used the relation
$
\lim_{\tau\to\infty}\rho_\tau(\eta)=\rho_\mathrm{st}(\eta)
.
$
In Eq. (\ref{eq:rhoTauSex}), the factor $\rho_\mathrm{can}(\eta)$ disappears
 because 
$
\lavetrajs\hat{S}_\mathrm{I}\ravetrajs^{\mathrm{es}}_{\eta_\tau=\eta}
-\lavetrajs\hat{S}_\mathrm{I}\ravetrajs^{\mathrm{es}}_{\eta_0=\eta^*}
$
 is equal to
$
\lavetrajs\hat{S}\ravetrajs^{\mathrm{es}}_{\eta_\tau=\eta}
-\lavetrajs\hat{S}\ravetrajs^{\mathrm{es}}_{\eta_0=\eta^*}
+2\beta^\mathrm{eq}H(\eta)
$
 apart from $\eta$-independent parts.

\begin{figure}[h]
\begin{center}
\includegraphics[scale=0.3]{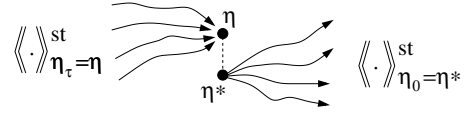}
\end{center}
\caption{Schematics for two kinds of CPEA.
Here $\eta^*$ is the time-reversed microstate of $\eta$.
The arrows show path of the system.
}
\label{fig:CEAschem}
\end{figure}

We further divide the contribution of the second term
 of the exponent on the r.h.s. in Eq. (\ref{eq:rhoTauSex})
 into two parts: one is that from equilibrium to steady state,
 and the other is that from the steady to the specified state $\eta$.
Because both the contributions converge within finite time,
 we decompose
$
\lavetrajs
{\hat S}_\mathrm{ex}
\ravetrajs^{\mathrm{es}}_{\eta_\tau=\eta}
$
into
$
\lavetrajs
{\hat S}_\mathrm{ex}
\ravetrajs^{\mathrm{st}}_{\eta_\tau=\eta}
+
\lavetrajs
{\hat S}_\mathrm{ex}
\ravetrajs^{\mathrm{es}}
$
for sufficiently large $\tau$.
We further omit the second term on r.h.s which does not depend on $\eta$.
From the definitions of the CPEAs, we have
$\lavetrajs
{\hat S}_\mathrm{ex}
\ravetrajs^{\mathrm{es}}_{\eta_0=\eta^*}
=
\lavetrajs
{\hat S}_\mathrm{ex}
\ravetrajs^{\mathrm{st}}_{\eta_0=\eta^*}
.
$
Then Eq. (\ref{eq:rhoTauSex}) leads to a neat expression
\begin{subequations}
\label{eq:rhost}
\begin{equation}
\rho_\mathrm{st}(\eta)
\propto
e^{{\Phi}(\eta)+O(|\bm{\epsilon}|^3)}
,
\end{equation}
\begin{equation}
{\Phi}(\eta):=
\lim_{\tau\rightarrow\infty}
\frac{1}{2}\left[
\lavetrajs
\hat{S}_\mathrm{ex}
\ravetrajs^{\mathrm{st}}_{\eta_\tau=\eta}
-
\lavetrajs
\hat{S}_\mathrm{ex}
\ravetrajs^{\mathrm{st}}_{\eta_0=\eta^*}
\right],
\end{equation}
\end{subequations}
which is our main result.
Note that ${\Phi}(\eta)$, which plays the role of 
 $-\beta H(\eta)$ in the equilibrium, is
 the difference between
 the excess entropy changes
 evaluated in the CPEAs 
 which have $\eta$ as the final state and
 $\eta^*$ as the initial state.
 (see Fig. \ref{fig:CEAschem}.)

It is quite interesting that such a combination
 of the first moments can give the result
 which is correct up to $O(|\bm{\epsilon}|^2)$.
It is crucial that these excess entropy changes depend only on
 transient regimes near the final or initial time.
This is in a sharp contrast between
 the formal expression (\ref{eq:RhoTauEsA}),
 which includes the average over the whole history.

It may be convenient, e.g. for actual measurements,
 to rewrite the excess entropy change $\hat{S}_\mathrm{ex}$
 in terms of heat transfer.
Using the definition (\ref{eq:Sdef}),
 Eq. (\ref{eq:rhost}) can be written as
\begin{subequations}
\label{eq:SexHeat}
\begin{equation}
\rho_\mathrm{st}(\eta)
\propto
e^{{\Phi}(\eta)+O(|\bm{\epsilon}|^3)}
,\;\;
{\Phi}(\eta)=-\sum_k \beta_k \mathcal{E}_k(\eta)
\end{equation}
\begin{equation}
\mathcal{E}_k(\eta):=\lim_{\tau\rightarrow\infty}\frac{1}{2}
\left[
\lavetrajs
{\hat Q}_{\mathrm{ex},k}
\ravetrajs^{\mathrm{st}}_{\eta_\tau=\eta}
-
\lavetrajs
{\hat Q}_{\mathrm{ex},k}
\ravetrajs^{\mathrm{st}}_{\eta_0=\eta^{*}}
\right]
,
\label{eq:Ek}
\end{equation}
\end{subequations}
 where the excess heat transfer ${\hat{Q}}_{\mathrm{ex},k}$
 is defined as
\begin{equation}
{\hat{Q}}_{\mathrm{ex},k}:={\hat{Q}}_k-\tau{J}_k^\mathrm{st}.
\end{equation}
Here ${J}_k^\mathrm{st}$ is the steady heat flow at NESS, and one has
 $\sigma=-\sum_k \beta_k {J}_k^\mathrm{st}$.
Note that $\mathcal{E}_k(\eta)$ is essentially the difference
 between the excess heat transfers
 evaluated in the CPEAs 
 which have $\eta$ as the final state and
 $\eta^*$ as the initial state.
Since the ensemble averages of the excess heat transfer are supposed
 to show good convergence as in \cite{KomatsuNakagawa},
 the limit $\tau\rightarrow\infty$ is expected to be
 effectively realized for moderately large $\tau$.

When the system in the equilibrium state
 at the inverse temperature $\beta$,
 we have the relation
$\bm{\beta}\cdot\mathcal{E}(\eta)=\beta\sum_k \mathcal{E}_k(\eta)
	=\beta [H(\eta)-\langle{H}\rangle^\mathrm{eq}]
$,
 where $\langle{H}\rangle^\mathrm{eq}$
 is the mean energy of the system at the equilibrium.
Thus the Boltzmann factor $\exp(-\beta H(\eta))$ is, of course,
 included in the form of Eqs. (\ref{eq:SexHeat}) as a special case.

\paragraph*{Numerical demonstrations:}
In order to demonstrate performance of our expression,
 we here show the results for a thermally driven ratchet
 \cite{Sekimoto,KomatsuNakagawa}
 where Langevin description for the heat baths is adopted.
The model consists of one translational degree of freedom $x$ and another degree $y$.
The degrees $x$ and $y$ are
 coupled to heat baths with inverse temperatures $\beta_x$ and $\beta_y$, respectively.
The system has periodic structure in $x$ and the external force $f$ is applied on $x$.
The time evolution of the system is described
 by the set of Langevin equations,
\begin{equation}
\left\lbrace
\begin{array}{rcl}
\dot{x} &=& \sqrt{2/\beta_x}\,{\xi_x}(t)
-{\partial U(x,y)}/{\partial x}+f,\\
\dot{y} &=& \sqrt{2/\beta_y}\,{\xi_y}(t)
-{\partial U(x,{y})}/{\partial{{y}}},
\end{array}
\right.
\label{eq:model}
\end{equation}
 where
 $U(x,y)=\exp(-y+\phi(x))+y^2/2,\;$
 $\phi(x)=-\sin(2\pi x)/2-\sin(4\pi x)/12+1/2$
 and $\xi_{*}(t)$ represent Gaussian white noises
 with a variance of unity.
For a reference of energy scale,
 the energy barrier of this ratchet potential is about $0.81$.

\begin{figure}
\begin{center}
\includegraphics[scale=0.5]{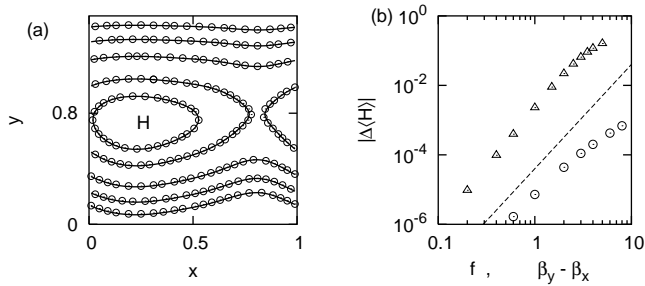}
\end{center}
\caption{Verification of the expression (\ref{eq:SexHeat})
 for the stationary distribution $\rho_\mathrm{st}(\eta)$ in NESS.
(a) The contours of ${\Phi}(\eta)=-\bm{\beta}\cdot\mathcal{E}(\eta)$ (circles)
 and $\ln \rho_\mathrm{st}(\eta)$ (solid lines)
 are plotted for $(\beta_x,\beta_y,f)=(1,10,0.8)$.
Eqs. (\ref{eq:SexHeat}) seems to work rather well
 even though $\beta_y/\beta_x=10$.
Contours are drawn at \{0.5,0,-1,-2,-3\}
 and most probable states are marked with ``H''.
The values of ${\Phi}(\eta)$ are shifted
 with some constant values which correspond to a normalization constant.
(b) Error estimation by the difference
 $|\Delta\langle H\rangle|:=|\sum_{\eta} H(\eta) [\rho_\mathrm{st}(\eta)-\exp({\Phi}(\eta))/Z]|$
 where $Z$ is a normalization constant.
Triangles show data set varying $f$ with $\beta_x=\beta_y=2$.
Circles show data set varying $\beta_y$ with $\beta_x=2,f=0$.
The drawn guide line is $\epsilon^{3}$.
For the calculation of $\rho_\mathrm{st}(\eta)$, we numerically solved
 Focker-Planck equations corresponding to
 the Langevin equations (\ref{eq:model}).
For ${\Phi}(\eta)$, we constructed and solved
 a set of difference equations for CPEAs
 of the excess heat transfer.
We have checked these results are consistent with
 results obtained by direct simulation of Langevin equations
 with stochastic energetics \cite{Sekimoto}.
}
\label{fig:demo}
\end{figure}

We measured ${\Phi}(\eta)$ by observing excess heat transfer
 from each heat bath.
In Fig. \ref{fig:demo}(a), comparison between
 ${\Phi}(\eta)=-\bm{\beta}\cdot\mathcal{E}(\eta)$
 and $\ln \rho_\mathrm{st}(\eta)$
 is shown for the cases in which both temperature difference and external field
 are simultaneously applied to the system.
We can see the expression works well
 even when the temperature difference is large.
As a detection for deviations of the expression (\ref{eq:SexHeat})
 from the true $\rho_\mathrm{st}(\eta)$,
 Fig. \ref{fig:demo}(b) shows the differences
 between expectation value of energy averaged with $\rho_\mathrm{st}(\eta)$
  and that with $\exp[{\Phi}(\eta)]/Z$ where $Z$ is a normalization constant.
From the figure,
 the error is estimated as $O(|\bm{\epsilon}|^3)$.
This clearly supports our theoretical result.

\paragraph*{Discussions:}

It is worth pointing out that in the phenomenological approach
 to NESS developed in \cite{OonoPaniconi} (See also \cite{HatanoSasa}),
 ``excess'' heat transfers rather than ``bare'' heat transfers 
 play fundamental roles.
The fact that our novel expressions (\ref{eq:rhost}), (\ref{eq:SexHeat})
 are also based on excess heat transfers
 is quite suggestive and encouraging.
A challenging future problem is to investigate
 characterizations of NESS based on Eqs. (\ref{eq:SexHeat}) or (\ref{eq:rhost}),
 e.g. stability of NESS, transition between NESSs,
 the principles of heat transfer enhancement
 previously suggested in \cite{KomatsuNakagawa}, and so on.

It is also interesting to confirm the expression Eqs. (\ref{eq:SexHeat})
 by precisely measuring heat transfers in nonequilibrium steady states
 in real systems.
Although such experiments need rather high skills
 to sustain temperature difference in a small system
 and to measure heat transfer with a sufficiently high resolution in time,
 we believe such measurements are possible,
 for instance in the mesoscopic low temperature system \cite{Pekola}.

In summary, the expression presented here
 is a natural extension of the canonical distribution.
In NESS, the energy of the microstate $\eta$ is not sufficient
 to specify the probability.
We need more precise information,
 which turns out to be conditioned path ensemble averages of the excess heat transfer.
One is the average for paths reaching $\eta$
 and the other is that for paths leaving the time-reversed microstate $\eta^*$.
In other words, the probability is determined by the difference
 between the excess entropy change until reaching $\eta$
 and that after leaving $\eta^*$.
In equilibrium, one-way information is sufficient
 because of the symmetry between the two,
 but in nonequilibrium the combination of the two becomes essential.
The expression presented here
 is correct up to the second order in thermodynamic force.
Higher order corrections can be obtained in a straightforward manner.
We hope researches following the present results will open
 a route to new concepts for the structure embedded in NESS.

We are very grateful to S. Sasa and H. Tasaki for discussions
 and critical readings of this manuscript,
 and indebted to H. Tasaki for simplifying
 an earlier version of the derivation.


\begin{thebibliography}{999}

\bibitem{Kubo}
{R. Kubo, M. Toda and N. Hashitsume}:
{\it Statistical Physics II: Noneqilibrium Stastical Mechanics}
{(Springer-Verlag, Berlin, 1991)}.

\bibitem{Zubarev} 
{D. N. Zubarev}:
{\it Nonequilibrium Statistical Thermodynamics}
{(Consultants Bureau, New York, 1974)}.

\bibitem{McLennan}
{J. A. McLennan}:
{\it Introduction to Nonequilibrium Statistical Mechanics}
{(Prentice Hall, Englewood Cliffs, NJ, 1990)}.

\bibitem{KawasakiGunton}
{K. Kawasaki and J. D. Gunton}:
{Phys. Rev. A}
\textbf{8}
{(1973)}
{2048}.

\bibitem{FT:Evans} 
{D. J. Evans, E. G. D. Cohen, and G. P. Morris}:
{Phys. Rev. Lett.}
\textbf{71}
{(1993)}
{2401}.

\bibitem{FT:Gallavotti} 
{G. Gallavotti and E. G. D. Cohen}:
{Phys. Rev. Lett.}
\textbf{74}
{(1995)}
{2694}.

\bibitem{FT:Kurchan}
{J. Kurchan}:
{J. Phys. A: Math. Gen.}
\textbf{31}
{(1998)}
{3719}.

\bibitem{FT:Maes} 
{C. Maes}:
{J. Stat. Phys.}
\textbf{95}
{(1999)}
{367}.

\bibitem{FT:Crooks} 
{G. E. Crooks}:
{Phys. Rev. E}
\textbf{61}
{(2000)}
{2361}.

\bibitem{Jarzynski}
{C. Jarzynski}:
{J. Stat. Phys.}
\textbf{98}
{77}
{(2000)}.
Eq. (23).

\bibitem{OnsagerMachlup}
{L. Onsager}:
{Phys. Rev.}
\textbf{91}
{(1953)}
{1505}.

\bibitem{HayashiSasa}
{K. Hayashi and S. Sasa}:
{Physica A}
\textbf{370}
{(2006)}
{407}.

\bibitem{KomatsuNakagawa}
{T. S. Komatsu and N. Nakagawa}:
{Phys. Rev. E}
\textbf{73}
{(2006)}
{065107(R)}.

\bibitem{Sekimoto}
{K. Sekimoto}:
{J. Phys. Soc. Jpn.}
\textbf{66}
{(1997)}
{1234}.

\bibitem{OonoPaniconi}
{Y. Oono and M. Paniconi}:
{Prog. Theor. Phys. Suppl.}
\textbf{130}
{(1998)}
29.

\bibitem{HatanoSasa} 
{T. Hatano and S. Sasa}:
{Phys. Rev. Lett.}
\textbf{86}
{(2001)}
{3463}.

\bibitem{Pekola} 
{F. Giazotto, T.T. Heikkil, A. Luukanen, A.M. Savin, and J.P.Pekola}:
{Rev. Mod. Phys.}
\textbf{78}
{(2006)}
{217}.

\end{thebibliography}
\end{document}